\documentclass[aps,amsmath,amssymb,physrev,reprint,superscriptaddress]{revtex4-2}

\usepackage{siunitx}
\usepackage{graphicx}
\usepackage[normalem]{ulem}
\usepackage{xcolor}
\usepackage{multirow}
\usepackage{float}
\begin{document}

\author{Josh T. Christensen}
\author{Farhan Azeem}
\author{Luke S. Trainor}
\author{Dmitry V. Strekalov}
\affiliation{The Dodd-Walls Centre for Photonic and Quantum Technologies, New Zealand}
\affiliation{Department of Physics, University of Otago, 730 Cumberland Street, Dunedin 9016, New Zealand}
\author{Harald G. L. Schwefel}
\email{harald.schwefel@otago.ac.nz}
\affiliation{The Dodd-Walls Centre for Photonic and Quantum Technologies, New Zealand}
\affiliation{Department of Physics, University of Otago, 730 Cumberland Street, Dunedin 9016, New Zealand}

\title{Distance calibration via Newton's rings in yttrium lithium fluoride whispering gallery mode resonators}

\date{\today}

\begin{abstract}
In this work, we analyze the first whispering gallery mode resonator (WGMR) made from monocrystalline yttrium lithium fluoride (YLF). The disc-shaped resonator is fabricated using single-point diamond turning and exhibits a high intrinsic quality factor ($Q$) on the order of 10\textsuperscript{9}. Moreover, we employ a novel method based on microscopic imaging of Newton's rings through the back of a trapezoidal prism. This method can be used to evanescently couple light in to a WGMR and monitor the separation between the cavity and the coupling prism. Accurately calibrating the distance between a coupling prism and a WGMR is desirable as it can be used to improve experimental control and conditions, i.e., accurate coupler gap calibration can aid in tuning into desired coupling regimes and can be used to avoid potential damage caused by collisions between the coupling prism and the WGMR. Here, we use two different trapezoidal prisms together with the high-$Q$ YLF WGMR to demonstrate and discuss this method. 
\end{abstract}

\maketitle

\begin{figure*}
\centering
\includegraphics[width=\linewidth]{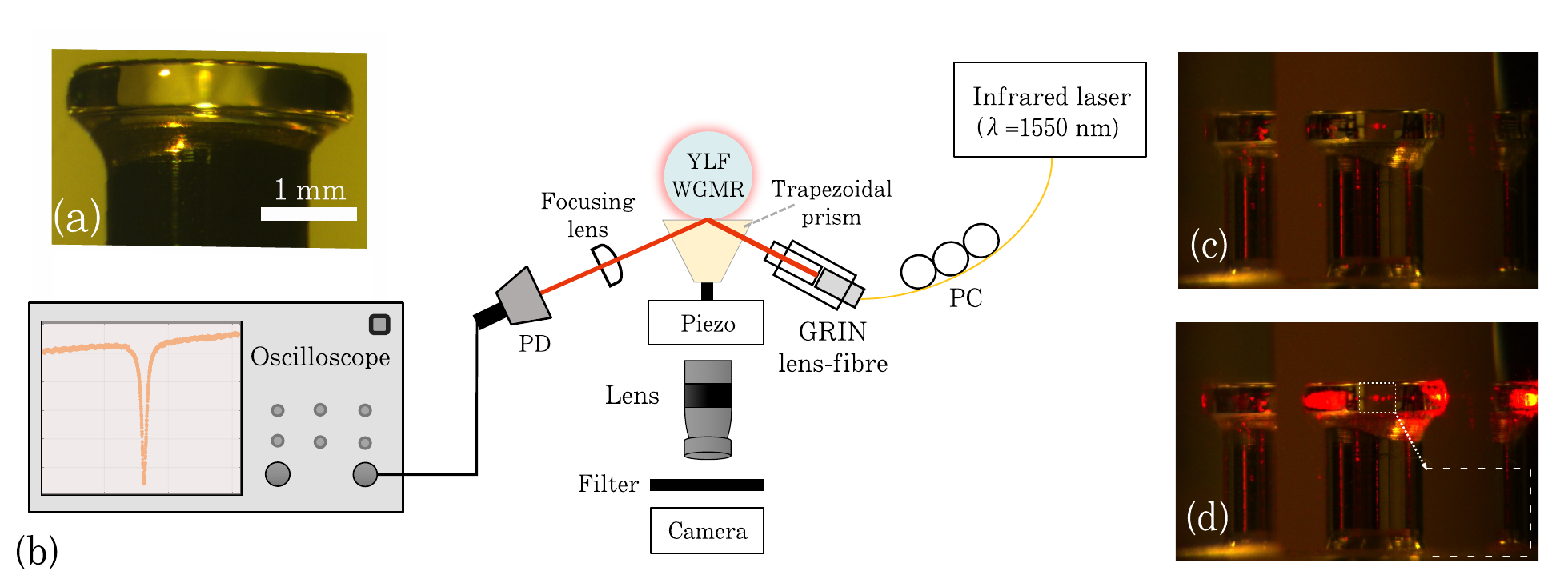}
\caption{\small A schematic explaining the experiment. (a) Polished YLF WGMR can be seen here after the fabrication process. (b) Shows the experimental setup used to couple light into the YLF WGMR using a trapezoidal prism. A camera is used to observe the interference pattern formed by the WGMR and the prism. A near-infrared laser is coupled in to the WGMR via a fiber pigtailed ferrule and a graded-index (GRIN) lens. The WGMs are observed using a PD and an oscilloscope.
A polarization controller (PC) is used to switch between TE and TM polarizations. A band-pass filter is later placed in front of the camera during data collection. (c) Shows an image of the YLF resonator seen from the back of the trapezoidal BK7 prism. The red light seen in the picture is a result of a red laser used for coupling light. (d) Image of the resonator when light is coupled inside it. Inset shows the contact region between the WGMR and the prism when the distance between them is negligible. This can also be used to observe coupling, for more information see supplement.}\label{fig:YLFsetup}
\end{figure*}

Optical resonators can store electromagnetic radiation and enhance various nonlinear interactions. One key factor is the rate with which light is coupled into the resonator. In a Fabry–Pérot resonator the coupling is determined by the reflectivity of the mirrors and can not easily be adjusted if the losses within the resonator change, e.g.\ as some nonlinear process starts. Whispering gallery mode (WGM) resonators allow an elegant control of the coupling rates. A whispering gallery mode resonator (WGMR) consists of a convex transparent dielectric~\cite{strekalov_nonlinear_2016} usually shaped as a sphere~\cite{humayun_far-infrared_2018}, toroid~\cite{carmon_static_2008}, ring~\cite{lambert_ultra-stable_2021}, or disc~\cite{norman_measuring_2022,azeem_ultralow_2022}, where the light is trapped via total internal reflection along the rim. Coupling is achieved through the evanescent field. By varying the distance of the coupler e.g.\ a prism~\cite{trainor_selective_2018} or a tapered fiber~\cite{qureshi_soliton_2022}  the coupling rate can be adjusted. Knowledge of the exact distance between the coupler and the WGMR is difficult to acquire to the $\sim\SI{10}{\nano\meter}$ precision required. Here, we present a new method that provides this distance by observing the interference pattern formed between the rim of the resonator and the coupling prism surface as observed through the back side of a trapezoidal prism (see Fig.~\ref{fig:YLFsetup}(d)). These, so called Newton's rings, allow for an absolute calibration of the distance between the prism and the resonator. We compare them with an independent measurement of the resonator linewidth.

We fabricate our resonator out of yttrium lithium fluoride (YLF) which is a uniaxial crystal and is commonly used as solid-state laser material as it has a small nonlinearity~\cite{brown_application_2016}. 
The yttrium ion can easily be exchanged with various rare earth ions~\cite{demirbas_comparative_2021}---such as holmium~\cite{hemmati_efficient_1987}, neodymium~\cite{maker_mode_1989}, erbium~\cite{schmaul_er_1993}, thulium~\cite{strauss_tmylf_2012}, praseodymium~\cite{luo_direct_2019} and ytterbium~\cite{demirbas_comparative_2021}--- with little deformation of the crystal structure, and leading to active materials for lasing and amplification.
YLF is transparent at ultraviolet wavelengths~\cite{yanagida_crystal_2009} and it exhibits anomalous dispersion in the telecom regime, making it a potential candidate for Kerr frequency comb generation~\cite{kippenberg_microresonator-based_2011}.
Finally, YLF combines a positive coefficient of thermal expansion with a negative thermorefractive coefficient at room temperature~\cite{aggarwal_measurement_2005}. This combination is known to lead to a wealth of thermo-optical effects \cite{he2009OscillatoryThermalDynamics} that could be used for optical comb generation, and stabilization \cite{kobatake2016Thermal}. We do observe such effects in our resonator and plan to report this research elsewhere.
Here, we report the first WGMR fabricated from monocrystalline YLF to the best of our knowledge. The YLF WGMR was fabricated via shaping an $a$-cut (see supplement) YLF crystal into a disc, using the single point diamond turning technique~\cite{sedlmeir2016crystalline}. Once the resonator was shaped, it was polished with diamond solutions of varying grain sizes, i.e., \SI{9}{\micro\meter} to  \SI{1}{\micro\meter}, in order to improve the surface quality. The fabricated YLF WGMR can be seen in Fig.~\ref{fig:YLFsetup}(a). The finalized resonator had a major radius of \SI{1.47+-0.02}{\milli\meter} and a minor radius of \SI{0.72+-0.01}{\milli\meter}.
After the fabrication, the resonator was cleaned and transferred to an experimental setup, shown in Fig.~\ref{fig:YLFsetup}(b). 
The YLF WGMR exhibits a high intrinsic $Q$-factor of $10^9$. 

We used an isosceles trapezoidal shaped prism to evanescently couple to the resonator. The benefit of this shape is that it allows a visible access port through the short parallel face to observe the Newton's rings formed at the coupling spot, when illuminated through the microscope and observed in reflection (see Fig.~\ref{fig:YLFsetup}(d)). This allowed us also to easily couple to the resonator by using a red guiding laser.
Newton's rings were observed using a lens and a camera and as shown in Fig.~\ref{fig:YLFsetup}(c) and (d). Fig.~\ref{fig:YLFsetup}(c) was captured when there was a large gap between the WGMR and the prism, whereas Fig.~\ref{fig:YLFsetup}(d) was taken when there was a minimal gap between the resonator and the prism and coupling was observed, which can be seen in the form of the resonator glowing, particularly on the right-hand side due to the unidirectional modes excited in the resonator. Insets in Fig.~\ref{fig:YLFsetup}(d) show the contact region between the WGMR and the prism, showing the observed Newton's rings. For more on the coupling process via the trapezoidal prism refer to the supplemental document.

After observing coupling with the guiding red laser, it was substituted with a tunable near-infrared (near-IR) laser with a central wavelength of \SI{1550}{\nano\meter} to observe WGMs. The trapezoidal prism was connected to a piezo stage to enable movement of the prism towards or away from the resonator through a voltage change. 

In order to use the Newton's rings for a precise measurement of distance, a narrow band-pass filter with a central wavelength of $\lambda_f=\SI{532}{\nm}$ was added to the setup. After the laser light was coupled into the WGMR, the output was collected through a focusing lens and a photodiode (PD). The PD was connected to an oscilloscope to observe the WGMs. The camera and the oscilloscope were connected to a computer for recording the images of the Newton's rings and the WGM spectra.
\begin{figure*}
\centering
\includegraphics[width=1\linewidth]{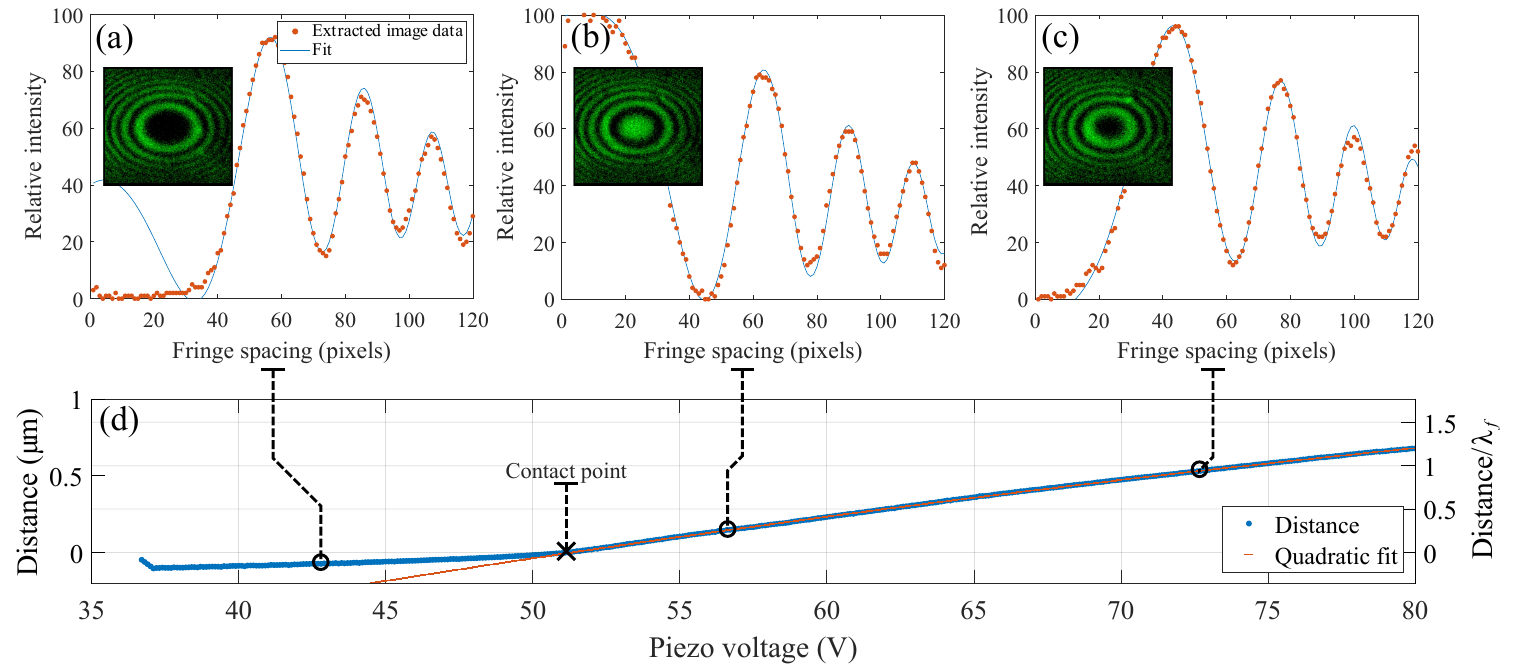}
\caption{\small The data extracted from the Newton's rings (shown in insets) observed using the camera through the trapezoidal prism, with a green band-pass filter. (a) 
Shows the relative intensity data when the prism and WGMR are in contact. The colour of the center becomes dark when this happens and it can be observed that it has expanded, which indicates that the prism is squeezing the WGMR, this dark center is not a Newton's ring so \eqref{eq:sinfit} no longer describes it.
(b) At a distance of about $\lambda_f/4$ constructive interference leads to a bright center (shown in the inset), so the relative intensity begins at a maximum.
In (c) -- at a distance of about $\lambda_f$ -- however, the center is dark due to destructive interference.
(d) Shows the effect of the piezo on the distance of separation with reference to the relative intensity sub-figures (a)-(c) at their respective piezo voltages.
Above \SI{52}{\volt}, the piezo response is typical for a decreasing applied voltage.
Below \SI{52}{\volt} the behaviour strays from the expected quadratic response as the prism and resonator are in contact and are only being deformed as (a) shows an enlarged dark spot.
}\label{fig:YLFNewton}
\end{figure*}

Once the experiment was set up, high-$Q$ transverse electric (TE) and transverse magnetic (TM) polarized modes were chosen to study the behaviour of Newton's rings. In this paper we label modes that are mainly polarized normal to the resonator plane as TE polarized, which is not a universal choice. WGM data was recorded by moving the prism towards the resonator via the piezo stage from the under-coupled regime to extremely over-coupled, i.e., when the WGMR and the prism were in contact. This was achieved by systematically varying the piezo step voltage in increments of \SI{0.1}{\volt}, while an image of the Newton's rings was simultaneously acquired for each iteration via the camera. 
The colour images of the Newton's rings were converted to grayscale images and fed into a Python image analysis code. The code works by averaging the image data along elliptical trajectories. The centre of the ellipse, aspect ratio, and orientation are initially fixed, and the large axis length is varied. The oscillation amplitude of the resulting one-dimensional intensity distribution indicates how well the initial assumption matches the observed fringes. The code then varies these three parameters and repeats the elliptical averaging until the contrast maximum is reached, and returns their final values. Such optimized curves are shown in Fig.~\ref{fig:YLFNewton}(a)-(c). The advantage of this approach over taking a vertical or horizontal slice across the image is that every image pixel is used, and also the possible tilt of the fringes is accounted for.
Figures~\ref{fig:YLFNewton}(b) and (c) show how the pattern of Newton's rings changes when the gap between the WGMR and the prism varies.
In order to calibrate the distance we fit these patterns with
\begin{equation}
\label{eq:sinfit}
    I = p_1\sin^2{(p_2x^2 + \phi)}e^{-p_3x^2} +p_4+p_5x+p_6x^2,
\end{equation}
where $I$ is the relative intensity, and $p_i$, $\phi$ are the fit parameters, $p_i$ are discussed in the supplement. The phase $\phi$ is extracted and the process repeated at each voltage step.
The $\pi$ phase shift between alternate maxima in \eqref{eq:sinfit} corresponds to a half wavelength change in distance between prism and resonator and allows us to determine the separation distance through $d  = \phi\lambda_f / (2\pi),$
where $\lambda_f$ is the wavelength of light passed through the filter. 
When the prism touches the WGMR a large central dark spot forms, see Fig.~\ref{fig:YLFNewton}(a). This is attributed to the prism and the WGMR deforming in response to the applied pressure and extending the contact area. Identifying the touching point is not straight forward from these figures as there is no significant change in the intensity data.
However, the characteristic enlarging of the center node allows us to find the absolute phase. 
A quadratic fit as shown in Fig.~\ref{fig:YLFNewton}(d) of the resulting distances as a function of voltage on the coupling piezo allows us to clearly identify the contact point.
We now study the effect of the coupler gap to properties of WGMs.  For this at each piezo step the laser was scanned over a resonance and the mode's linewidth was found by fitting the reflection to a Lorentzian function.
An exponential fit was applied to the fitted linewidths as a function of coupling distance,
\begin{equation}
\label{eq:finalfit}
    \Delta\nu(d) = \Delta\nu_{c,0}e^{-2\kappa d} + \Delta\nu_i.
\end{equation}
This fit was used to determine the intrinsic linewidth $\Delta\nu_i$, the coupling linewidth at zero distance $\Delta\nu_{c,0}$, and the evanescent-field decay constant $\kappa$. The intrinsic linewidth is readily found from the measurements at large distances $d$, where the line shape is almost independent of the coupling. 
The evanescent-field decay constant is given theoretically by~\cite{gorodetsky_optical_1999}:
\begin{equation}
    \kappa = \frac{2 \pi}{\lambda_l} \sqrt{n^2_\text{res} - n^2_\text{out}}, \label{eq:kappa}
\end{equation}
where $n_\text{out}$ is the refractive index of the outside medium, which is air in our experiment, therefore $n_\text{out}\approx 1$. Note that the decay constant is the same for the evanescent field of both the resonator and prism as the coupling is phase matched.

\begin{table}
\centering
\small
\caption{Average evanescent-field decay constants and coupling linewidths for TE and TM polarizations with GGG and BK7 coupling prisms, when averaging over $N$ measurements of modes. All uncertainties are one standard deviation.}
\begin{tabular}{ c c c c} 
\hline
Prism & Parameter & TE & TM \\
\hline
\multirow{4}{4em}{GGG}& $\kappa$ (\si{\per\micro\meter}) & \num{4.03\pm0.33} & \num{4.23\pm0.18} \\ 
& $\Delta\nu_{c,0}$ (\si{\mega\hertz}) & \num{49\pm24} & \num{29\pm10}\\
& $\Delta\nu_i$ (\si{\kilo\hertz}) & \num{401\pm235}  & \num{677\pm131}\\ 
& $N$ & 20 & 6 \\ 
\hline
\multirow{4}{4em}{BK7} & $\kappa$ (\si{\per\micro\meter}) & \num{4.46\pm0.25}  & \num{5.52\pm0.65}\\ 
& $\Delta\nu_{c,0}$ (\si{\mega\hertz}) & \num{52\pm17}  & \num{16\pm3}\\ 
& $\Delta\nu_i$ (\si{\kilo\hertz}) & \num{186\pm20}  & \num{675\pm151}\\ 
& $N$ & 5  & 8\\ 
\hline
\end{tabular}
\label{table:1}
\end{table}

We have used two different prism materials in our study, BK7 glass ($n=1.50$) and gadolinium gallium garnet (GGG) $(n=1.94)$, as we expect to see a variation of the coupling rate at zero distance. 
We fit the measured linewidth as a function of distance to \eqref{eq:finalfit} and report the values in Table.~\ref{table:1}. We find that the intrinsic linewidths of TM polarized mode families are greater than that of TE polarized modes, as we expect for high-$Q$ resonators~\cite{sedlmeir_high-q_2014}.

\begin{figure}
\centering
\includegraphics[width=\linewidth]{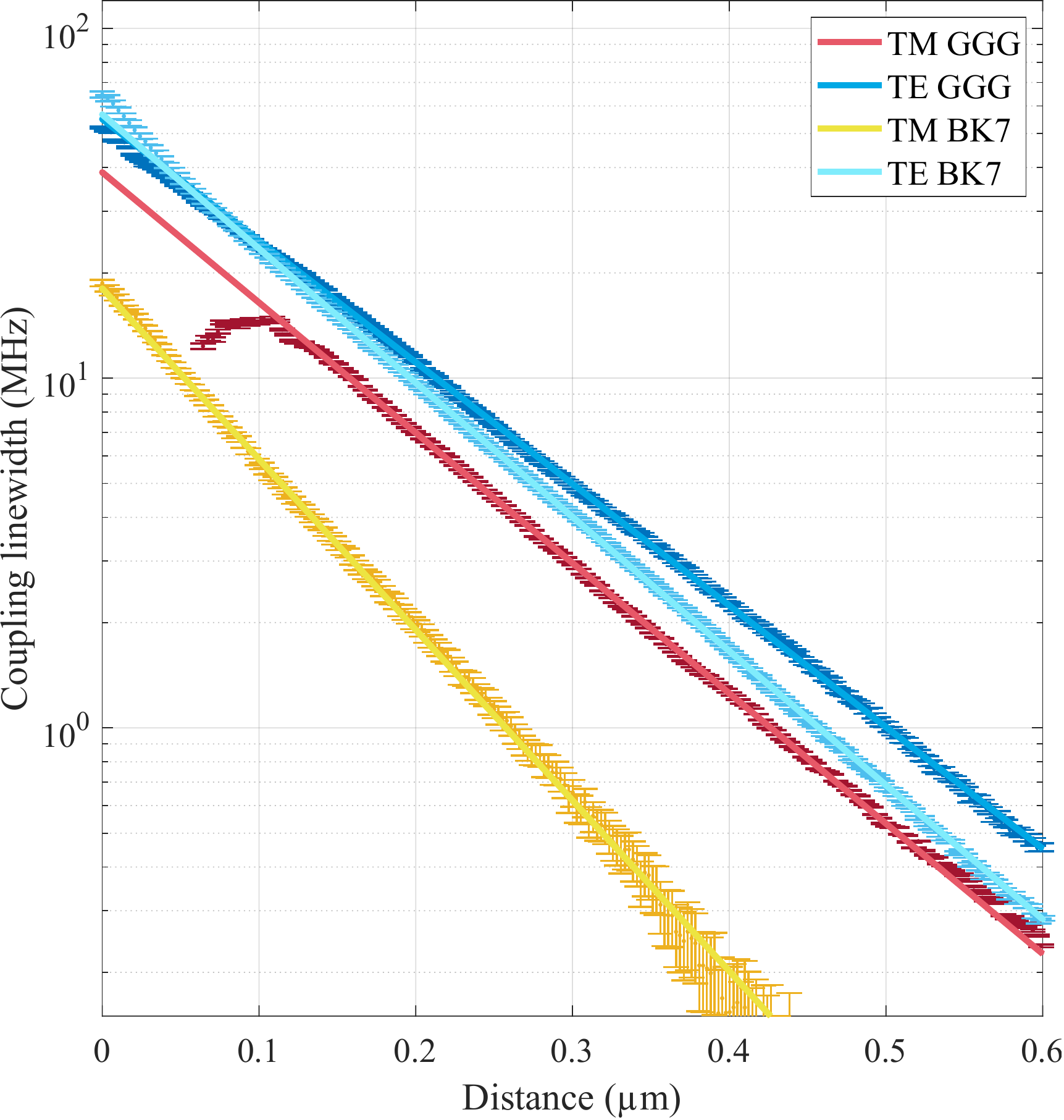}
\caption{\small
Coupling linewidth versus distance for  TE and TM polarized modes excited in the YLF WGMR using two different prisms, i.e., GGG and BK7. The respective intrinsic linewidth has been removed for each measurement. The distance was acquired from the Newton's rings data for each data point of linewidth. A fit applied to the data can be seen; used to verify the decay coefficient ($\kappa$) and coupling linewidth at zero distance ($\Delta\nu_{c,0}$) experimentally.
At small distances coupling arises to many parasitic modes which can distort the line shape and cause fitted linewidths to be inaccurate. 
}\label{fig:YLFdist}
\end{figure}

Removing the intrinsic linewidth from the measured linewidth, we find the coupling linewidth, which is shown in Fig.~\ref{fig:YLFdist}. The slope yields the decay constant $\kappa$ noted in Table.~\ref{table:1}. From \eqref{eq:kappa} we expected $\kappa$ of \SI{4.23}{\per\micro\meter} for TE polarized modes, that are oriented along the ordinary refractive index, and from \SI{4.23}{\per\micro\meter} to \SI{4.35}{\per\micro\meter} for TM polarized modes that vary from ordinary to extraordinary refractive indices depending on where on the resonator we are coupled. We do indeed find that a majority of the experimental values agree with the theory, but for TM modes with the BK7 prism we find a slight disparity.

Exponential fitting in Fig.~\ref{fig:YLFdist} also allows us to also find the coupling linewidth at zero distance $\Delta\nu_{c,0}$, shown in Table.~\ref{table:1}. 
This is when the accurate determination of the contact point $d=0$ is required, as the uncertainty in $d$ translates to the uncertainty in $\Delta\nu_{c,0}$. Determining the maximal coupling linewidth is very challenging in large resonators with dense mode spectra as adjacent modes start to overlap when highly over-coupled. Here, however, we have certainty of the contact point through the Newton's ring method described above.
The average TE linewidth is reasonably consistent with the theoretical estimate of 39 MHz found from both \cite{gorodetsky_optical_1999} and  \cite{gorodetsky1994coupling} for our coupler and resonator parameters.
It is generally acknowledged that the TM coupling linewidth is smaller than the TE coupling linewidth \cite{Savchenkov14Esens}, however sources differ on the amount by which the two polarizations differ \cite{savchenkov2018Q,foreman_dielectric_2016}.
For our resonator and a GGG prism, we find $\Delta\nu_{c,0}^{\rm TE}/\Delta\nu_{c,0}^{\rm TM}\approx1.7$ and for BK7 $\approx3.3$.
However, due to the large errors, these factors require more investigation. Furthermore, we have averaged over many mode families, rather then only examining fundamental modes.
This investigation is interesting because it will allow one to probe into the subtle aspects of the evanescent field coupling, in particular the back action of the coupler on the WGM field. 

In conclusion, we have demonstrated a high intrinsic $Q$-factor of $10^9$ in the first YLF WGMR evanescently coupled to a trapezoidal prism. We measure the geometric distance between the prism and the WGMR by directly observing the Newton's rings. Then we combine these results with measured coupling rates, which rely on the exact electromagnetic field distribution for a particular whispering gallery mode and its interplay with the coupler. A more detailed study with careful distinction of different mode families might shed light on the subtleties of unwanted tilts in the system.
In our study, we needed to make the resonator and prism touch in order to remove any distance ambiguity. This limitation could be removed by bichromatic imaging, e.g. by using a dual-bandpass filter and a color image sensor.\\ \\

\textbf{Data Availability.} \\ Data underlying the results presented in this paper are not publicly available at this time but may be obtained from the authors upon reasonable request.

\textbf{Funding} \\
D.V.S. and H.G.L.S. would like to thank the Ministry of Business, Innovation and Employment, New Zealand Catalyst Leaders fellowship (20-UOO-001-ILF).

\textbf{Disclosures} \\
The authors declare no conflicts of interest. 

\textbf{Supplemental document} \\
See Supplement for supporting content.


\begin{thebibliography}{28}%
\makeatletter
\providecommand \@ifxundefined [1]{%
 \@ifx{#1\undefined}
}%
\providecommand \@ifnum [1]{%
 \ifnum #1\expandafter \@firstoftwo
 \else \expandafter \@secondoftwo
 \fi
}%
\providecommand \@ifx [1]{%
 \ifx #1\expandafter \@firstoftwo
 \else \expandafter \@secondoftwo
 \fi
}%
\providecommand \natexlab [1]{#1}%
\providecommand \enquote  [1]{``#1''}%
\providecommand \bibnamefont  [1]{#1}%
\providecommand \bibfnamefont [1]{#1}%
\providecommand \citenamefont [1]{#1}%
\providecommand \href@noop [0]{\@secondoftwo}%
\providecommand \href [0]{\begingroup \@sanitize@url \@href}%
\providecommand \@href[1]{\@@startlink{#1}\@@href}%
\providecommand \@@href[1]{\endgroup#1\@@endlink}%
\providecommand \@sanitize@url [0]{\catcode `\\12\catcode `\$12\catcode
  `\&12\catcode `\#12\catcode `\^12\catcode `\_12\catcode `\%12\relax}%
\providecommand \@@startlink[1]{}%
\providecommand \@@endlink[0]{}%
\providecommand \url  [0]{\begingroup\@sanitize@url \@url }%
\providecommand \@url [1]{\endgroup\@href {#1}{\urlprefix }}%
\providecommand \urlprefix  [0]{URL }%
\providecommand \Eprint [0]{\href }%
\providecommand \doibase [0]{https://doi.org/}%
\providecommand \selectlanguage [0]{\@gobble}%
\providecommand \bibinfo  [0]{\@secondoftwo}%
\providecommand \bibfield  [0]{\@secondoftwo}%
\providecommand \translation [1]{[#1]}%
\providecommand \BibitemOpen [0]{}%
\providecommand \bibitemStop [0]{}%
\providecommand \bibitemNoStop [0]{.\EOS\space}%
\providecommand \EOS [0]{\spacefactor3000\relax}%
\providecommand \BibitemShut  [1]{\csname bibitem#1\endcsname}%
\let\auto@bib@innerbib\@empty
\bibitem [{\citenamefont {Strekalov}\ \emph {et~al.}(2016)\citenamefont
  {Strekalov}, \citenamefont {Marquardt}, \citenamefont {Matsko}, \citenamefont
  {Schwefel},\ and\ \citenamefont {Leuchs}}]{strekalov_nonlinear_2016}%
  \BibitemOpen
  \bibfield  {author} {\bibinfo {author} {\bibfnamefont {D.~V.}\ \bibnamefont
  {Strekalov}}, \bibinfo {author} {\bibfnamefont {C.}~\bibnamefont
  {Marquardt}}, \bibinfo {author} {\bibfnamefont {A.~B.}\ \bibnamefont
  {Matsko}}, \bibinfo {author} {\bibfnamefont {H.~G.~L.}\ \bibnamefont
  {Schwefel}},\ and\ \bibinfo {author} {\bibfnamefont {G.}~\bibnamefont
  {Leuchs}},\ }\bibfield  {title} {\bibinfo {title} {Nonlinear and quantum
  optics with whispering gallery resonators},\ }\href
  {https://doi.org/10.1088/2040-8978/18/12/123002} {\bibfield  {journal}
  {\bibinfo  {journal} {Journal of Optics}\ }\textbf {\bibinfo {volume} {18}},\
  \bibinfo {pages} {123002} (\bibinfo {year} {2016})}\BibitemShut {NoStop}%
\bibitem [{\citenamefont {Humayun}\ \emph {et~al.}(2018)\citenamefont
  {Humayun}, \citenamefont {Khan}, \citenamefont {Azeem}, \citenamefont
  {Chaudhry}, \citenamefont {Gökay}, \citenamefont {Murib},\ and\
  \citenamefont {Serpengüzel}}]{humayun_far-infrared_2018}%
  \BibitemOpen
  \bibfield  {author} {\bibinfo {author} {\bibfnamefont {M.~H.}\ \bibnamefont
  {Humayun}}, \bibinfo {author} {\bibfnamefont {I.}~\bibnamefont {Khan}},
  \bibinfo {author} {\bibfnamefont {F.}~\bibnamefont {Azeem}}, \bibinfo
  {author} {\bibfnamefont {M.~R.}\ \bibnamefont {Chaudhry}}, \bibinfo {author}
  {\bibfnamefont {U.~S.}\ \bibnamefont {Gökay}}, \bibinfo {author}
  {\bibfnamefont {M.~S.}\ \bibnamefont {Murib}},\ and\ \bibinfo {author}
  {\bibfnamefont {A.}~\bibnamefont {Serpengüzel}},\ }\bibfield  {title}
  {\bibinfo {title} {Far-infrared elastic scattering proposal for the
  {Avogadro} {Project}'s silicon spheres},\ }\href
  {https://doi.org/10.1016/j.jqsrt.2017.12.023} {\bibfield  {journal} {\bibinfo
   {journal} {Journal of Quantitative Spectroscopy and Radiative Transfer}\
  }\textbf {\bibinfo {volume} {210}},\ \bibinfo {pages} {173} (\bibinfo {year}
  {2018})}\BibitemShut {NoStop}%
\bibitem [{\citenamefont {Carmon}\ \emph {et~al.}(2008)\citenamefont {Carmon},
  \citenamefont {Schwefel}, \citenamefont {Yang}, \citenamefont {Oxborrow},
  \citenamefont {Stone},\ and\ \citenamefont {Vahala}}]{carmon_static_2008}%
  \BibitemOpen
  \bibfield  {author} {\bibinfo {author} {\bibfnamefont {T.}~\bibnamefont
  {Carmon}}, \bibinfo {author} {\bibfnamefont {H.~G.~L.}\ \bibnamefont
  {Schwefel}}, \bibinfo {author} {\bibfnamefont {L.}~\bibnamefont {Yang}},
  \bibinfo {author} {\bibfnamefont {M.}~\bibnamefont {Oxborrow}}, \bibinfo
  {author} {\bibfnamefont {A.~D.}\ \bibnamefont {Stone}},\ and\ \bibinfo
  {author} {\bibfnamefont {K.~J.}\ \bibnamefont {Vahala}},\ }\bibfield  {title}
  {\bibinfo {title} {Static {Envelope} {Patterns} in {Composite} {Resonances}
  {Generated} by {Level} {Crossing} in {Optical} {Toroidal} {Microcavities}},\
  }\href {https://doi.org/10.1103/PhysRevLett.100.103905} {\bibfield  {journal}
  {\bibinfo  {journal} {Physical Review Letters}\ }\textbf {\bibinfo {volume}
  {100}},\ \bibinfo {pages} {103905} (\bibinfo {year} {2008})}\BibitemShut
  {NoStop}%
\bibitem [{\citenamefont {Lambert}\ \emph {et~al.}(2021)\citenamefont
  {Lambert}, \citenamefont {Trainor},\ and\ \citenamefont
  {Schwefel}}]{lambert_ultra-stable_2021}%
  \BibitemOpen
  \bibfield  {author} {\bibinfo {author} {\bibfnamefont {N.~J.}\ \bibnamefont
  {Lambert}}, \bibinfo {author} {\bibfnamefont {L.~S.}\ \bibnamefont
  {Trainor}},\ and\ \bibinfo {author} {\bibfnamefont {H.~G.~L.}\ \bibnamefont
  {Schwefel}},\ }\bibfield  {title} {\bibinfo {title} {An ultra-stable
  microresonator-based electro-optic dual frequency comb},\ }\bibfield
  {journal} {\bibinfo  {journal} {arXiv preprint arXiv:2108.11140}\ }\href
  {https://doi.org/10.48550/ARXIV.2108.11140} {10.48550/ARXIV.2108.11140}
  (\bibinfo {year} {2021})\BibitemShut {NoStop}%
\bibitem [{\citenamefont {Norman}\ \emph {et~al.}(2022)\citenamefont {Norman},
  \citenamefont {Azeem}, \citenamefont {Longdell},\ and\ \citenamefont
  {Schwefel}}]{norman_measuring_2022}%
  \BibitemOpen
  \bibfield  {author} {\bibinfo {author} {\bibfnamefont {D.~S.}\ \bibnamefont
  {Norman}}, \bibinfo {author} {\bibfnamefont {F.}~\bibnamefont {Azeem}},
  \bibinfo {author} {\bibfnamefont {J.~J.}\ \bibnamefont {Longdell}},\ and\
  \bibinfo {author} {\bibfnamefont {H.~G.~L.}\ \bibnamefont {Schwefel}},\
  }\bibfield  {title} {\bibinfo {title} {Measuring optical loss in yttrium
  orthosilicate using a whispering gallery mode resonator},\ }\href
  {https://doi.org/10.1088/2633-4356/ac4c39} {\bibfield  {journal} {\bibinfo
  {journal} {Materials for Quantum Technology}\ }\textbf {\bibinfo {volume}
  {2}},\ \bibinfo {pages} {011001} (\bibinfo {year} {2022})}\BibitemShut
  {NoStop}%
\bibitem [{\citenamefont {Azeem}\ \emph {et~al.}(2022)\citenamefont {Azeem},
  \citenamefont {Trainor}, \citenamefont {Gao}, \citenamefont {Isarov},
  \citenamefont {Strekalov},\ and\ \citenamefont
  {Schwefel}}]{azeem_ultralow_2022}%
  \BibitemOpen
  \bibfield  {author} {\bibinfo {author} {\bibfnamefont {F.}~\bibnamefont
  {Azeem}}, \bibinfo {author} {\bibfnamefont {L.~S.}\ \bibnamefont {Trainor}},
  \bibinfo {author} {\bibfnamefont {A.}~\bibnamefont {Gao}}, \bibinfo {author}
  {\bibfnamefont {M.}~\bibnamefont {Isarov}}, \bibinfo {author} {\bibfnamefont
  {D.~V.}\ \bibnamefont {Strekalov}},\ and\ \bibinfo {author} {\bibfnamefont
  {H.~G.~L.}\ \bibnamefont {Schwefel}},\ }\bibfield  {title} {\bibinfo {title}
  {Ultra‐{Low} {Threshold} {Titanium}‐{Doped} {Sapphire}
  {Whispering}‐{Gallery} {Laser}},\ }\href
  {https://doi.org/10.1002/adom.202102137} {\bibfield  {journal} {\bibinfo
  {journal} {Advanced Optical Materials}\ }\textbf {\bibinfo {volume} {10}},\
  \bibinfo {pages} {2102137} (\bibinfo {year} {2022})}\BibitemShut {NoStop}%
\bibitem [{\citenamefont {Trainor}\ \emph {et~al.}(2018)\citenamefont
  {Trainor}, \citenamefont {Sedlmeir}, \citenamefont {Peuntinger},\ and\
  \citenamefont {Schwefel}}]{trainor_selective_2018}%
  \BibitemOpen
  \bibfield  {author} {\bibinfo {author} {\bibfnamefont {L.~S.}\ \bibnamefont
  {Trainor}}, \bibinfo {author} {\bibfnamefont {F.}~\bibnamefont {Sedlmeir}},
  \bibinfo {author} {\bibfnamefont {C.}~\bibnamefont {Peuntinger}},\ and\
  \bibinfo {author} {\bibfnamefont {H.~G.~L.}\ \bibnamefont {Schwefel}},\
  }\bibfield  {title} {\bibinfo {title} {Selective {Coupling} {Enhances}
  {Harmonic} {Generation} of {Whispering}-{Gallery} {Modes}},\ }\href
  {https://doi.org/10.1103/PhysRevApplied.9.024007} {\bibfield  {journal}
  {\bibinfo  {journal} {Physical Review Applied}\ }\textbf {\bibinfo {volume}
  {9}},\ \bibinfo {pages} {024007} (\bibinfo {year} {2018})}\BibitemShut
  {NoStop}%
\bibitem [{\citenamefont {Qureshi}\ \emph {et~al.}(2022)\citenamefont
  {Qureshi}, \citenamefont {Ng}, \citenamefont {Azeem}, \citenamefont
  {Trainor}, \citenamefont {Schwefel}, \citenamefont {Coen}, \citenamefont
  {Erkintalo},\ and\ \citenamefont {Murdoch}}]{qureshi_soliton_2022}%
  \BibitemOpen
  \bibfield  {author} {\bibinfo {author} {\bibfnamefont {P.~C.}\ \bibnamefont
  {Qureshi}}, \bibinfo {author} {\bibfnamefont {V.}~\bibnamefont {Ng}},
  \bibinfo {author} {\bibfnamefont {F.}~\bibnamefont {Azeem}}, \bibinfo
  {author} {\bibfnamefont {L.~S.}\ \bibnamefont {Trainor}}, \bibinfo {author}
  {\bibfnamefont {H.~G.~L.}\ \bibnamefont {Schwefel}}, \bibinfo {author}
  {\bibfnamefont {S.}~\bibnamefont {Coen}}, \bibinfo {author} {\bibfnamefont
  {M.}~\bibnamefont {Erkintalo}},\ and\ \bibinfo {author} {\bibfnamefont
  {S.~G.}\ \bibnamefont {Murdoch}},\ }\bibfield  {title} {\bibinfo {title}
  {Soliton linear-wave scattering in a {Kerr} microresonator},\ }\href
  {https://doi.org/10.1038/s42005-022-00903-5} {\bibfield  {journal} {\bibinfo
  {journal} {Communications Physics}\ }\textbf {\bibinfo {volume} {5}},\
  \bibinfo {pages} {123} (\bibinfo {year} {2022})}\BibitemShut {NoStop}%
\bibitem [{\citenamefont {Brown}\ \emph {et~al.}(2016)\citenamefont {Brown},
  \citenamefont {Tornegård}, \citenamefont {Kolis}, \citenamefont {McMillen},
  \citenamefont {Moore}, \citenamefont {Sanjeewa},\ and\ \citenamefont
  {Hancock}}]{brown_application_2016}%
  \BibitemOpen
  \bibfield  {author} {\bibinfo {author} {\bibfnamefont {D.}~\bibnamefont
  {Brown}}, \bibinfo {author} {\bibfnamefont {S.}~\bibnamefont {Tornegård}},
  \bibinfo {author} {\bibfnamefont {J.}~\bibnamefont {Kolis}}, \bibinfo
  {author} {\bibfnamefont {C.}~\bibnamefont {McMillen}}, \bibinfo {author}
  {\bibfnamefont {C.}~\bibnamefont {Moore}}, \bibinfo {author} {\bibfnamefont
  {L.}~\bibnamefont {Sanjeewa}},\ and\ \bibinfo {author} {\bibfnamefont
  {C.}~\bibnamefont {Hancock}},\ }\bibfield  {title} {\bibinfo {title} {The
  {Application} of {Cryogenic} {Laser} {Physics} to the {Development} of {High}
  {Average} {Power} {Ultra}-{Short} {Pulse} {Lasers}},\ }\href
  {https://doi.org/10.3390/app6010023} {\bibfield  {journal} {\bibinfo
  {journal} {Applied Sciences}\ }\textbf {\bibinfo {volume} {6}},\ \bibinfo
  {pages} {23} (\bibinfo {year} {2016})}\BibitemShut {NoStop}%
\bibitem [{\citenamefont {Demirbas}\ \emph {et~al.}(2021)\citenamefont
  {Demirbas}, \citenamefont {Kellert}, \citenamefont {Thesinga}, \citenamefont
  {Hua}, \citenamefont {Reuter}, \citenamefont {Kärtner},\ and\ \citenamefont
  {Pergament}}]{demirbas_comparative_2021}%
  \BibitemOpen
  \bibfield  {author} {\bibinfo {author} {\bibfnamefont {U.}~\bibnamefont
  {Demirbas}}, \bibinfo {author} {\bibfnamefont {M.}~\bibnamefont {Kellert}},
  \bibinfo {author} {\bibfnamefont {J.}~\bibnamefont {Thesinga}}, \bibinfo
  {author} {\bibfnamefont {Y.}~\bibnamefont {Hua}}, \bibinfo {author}
  {\bibfnamefont {S.}~\bibnamefont {Reuter}}, \bibinfo {author} {\bibfnamefont
  {F.~X.}\ \bibnamefont {Kärtner}},\ and\ \bibinfo {author} {\bibfnamefont
  {M.}~\bibnamefont {Pergament}},\ }\bibfield  {title} {\bibinfo {title}
  {Comparative investigation of lasing and amplification performance in
  cryogenic {Yb}:{YLF} systems},\ }\href
  {https://doi.org/10.1007/s00340-021-07588-8} {\bibfield  {journal} {\bibinfo
  {journal} {Applied Physics B}\ }\textbf {\bibinfo {volume} {127}},\ \bibinfo
  {pages} {46} (\bibinfo {year} {2021})}\BibitemShut {NoStop}%
\bibitem [{\citenamefont {Hemmati}(1987)}]{hemmati_efficient_1987}%
  \BibitemOpen
  \bibfield  {author} {\bibinfo {author} {\bibfnamefont {H.}~\bibnamefont
  {Hemmati}},\ }\bibfield  {title} {\bibinfo {title} {Efficient holmium:yttrium
  lithium fluoride laser longitudinally pumped by a semiconductor laser
  array},\ }\href {https://doi.org/10.1063/1.98348} {\bibfield  {journal}
  {\bibinfo  {journal} {Applied Physics Letters}\ }\textbf {\bibinfo {volume}
  {51}},\ \bibinfo {pages} {564} (\bibinfo {year} {1987})}\BibitemShut
  {NoStop}%
\bibitem [{\citenamefont {Maker}\ and\ \citenamefont
  {Ferguson}(1989)}]{maker_mode_1989}%
  \BibitemOpen
  \bibfield  {author} {\bibinfo {author} {\bibfnamefont {G.~T.}\ \bibnamefont
  {Maker}}\ and\ \bibinfo {author} {\bibfnamefont {A.~I.}\ \bibnamefont
  {Ferguson}},\ }\bibfield  {title} {\bibinfo {title} {Mode locking and
  \textit{{Q}} switching of a diode laser pumped neodymium‐doped yttrium
  lithium fluoride laser},\ }\href {https://doi.org/10.1063/1.100976}
  {\bibfield  {journal} {\bibinfo  {journal} {Applied Physics Letters}\
  }\textbf {\bibinfo {volume} {54}},\ \bibinfo {pages} {403} (\bibinfo {year}
  {1989})}\BibitemShut {NoStop}%
\bibitem [{\citenamefont {Schmaul}\ \emph {et~al.}(1993)\citenamefont
  {Schmaul}, \citenamefont {Huber}, \citenamefont {Clausen}, \citenamefont
  {Chai}, \citenamefont {LiKamWa},\ and\ \citenamefont
  {Bass}}]{schmaul_er_1993}%
  \BibitemOpen
  \bibfield  {author} {\bibinfo {author} {\bibfnamefont {B.}~\bibnamefont
  {Schmaul}}, \bibinfo {author} {\bibfnamefont {G.}~\bibnamefont {Huber}},
  \bibinfo {author} {\bibfnamefont {R.}~\bibnamefont {Clausen}}, \bibinfo
  {author} {\bibfnamefont {B.}~\bibnamefont {Chai}}, \bibinfo {author}
  {\bibfnamefont {P.}~\bibnamefont {LiKamWa}},\ and\ \bibinfo {author}
  {\bibfnamefont {M.}~\bibnamefont {Bass}},\ }\bibfield  {title} {\bibinfo
  {title} {Er $^{\textrm{3+}}$:YLiF$_{\textrm{4}}$ continuous wave cascade
  laser operation at 1620 and 2810 nm at room temperature},\ }\href
  {https://doi.org/10.1063/1.108904} {\bibfield  {journal} {\bibinfo  {journal}
  {Applied Physics Letters}\ }\textbf {\bibinfo {volume} {62}},\ \bibinfo
  {pages} {541} (\bibinfo {year} {1993})}\BibitemShut {NoStop}%
\bibitem [{\citenamefont {Strauss}\ \emph {et~al.}(2012)\citenamefont
  {Strauss}, \citenamefont {Esser}, \citenamefont {King},\ and\ \citenamefont
  {Maweza}}]{strauss_tmylf_2012}%
  \BibitemOpen
  \bibfield  {author} {\bibinfo {author} {\bibfnamefont {H.~J.}\ \bibnamefont
  {Strauss}}, \bibinfo {author} {\bibfnamefont {M.~J.~D.}\ \bibnamefont
  {Esser}}, \bibinfo {author} {\bibfnamefont {G.}~\bibnamefont {King}},\ and\
  \bibinfo {author} {\bibfnamefont {L.}~\bibnamefont {Maweza}},\ }\bibfield
  {title} {\bibinfo {title} {Tm:{YLF} slab wavelength-selected laser},\ }\href
  {https://doi.org/10.1364/OME.2.001165} {\bibfield  {journal} {\bibinfo
  {journal} {Optical Materials Express}\ }\textbf {\bibinfo {volume} {2}},\
  \bibinfo {pages} {1165} (\bibinfo {year} {2012})}\BibitemShut {NoStop}%
\bibitem [{\citenamefont {Luo}\ \emph {et~al.}(2019)\citenamefont {Luo},
  \citenamefont {Cai}, \citenamefont {Xu}, \citenamefont {Shen}, \citenamefont
  {Chen}, \citenamefont {Li},\ and\ \citenamefont {Cao}}]{luo_direct_2019}%
  \BibitemOpen
  \bibfield  {author} {\bibinfo {author} {\bibfnamefont {S.}~\bibnamefont
  {Luo}}, \bibinfo {author} {\bibfnamefont {Z.}~\bibnamefont {Cai}}, \bibinfo
  {author} {\bibfnamefont {H.}~\bibnamefont {Xu}}, \bibinfo {author}
  {\bibfnamefont {Z.}~\bibnamefont {Shen}}, \bibinfo {author} {\bibfnamefont
  {H.}~\bibnamefont {Chen}}, \bibinfo {author} {\bibfnamefont {L.}~\bibnamefont
  {Li}},\ and\ \bibinfo {author} {\bibfnamefont {Y.}~\bibnamefont {Cao}},\
  }\bibfield  {title} {\bibinfo {title} {Direct oscillation at 640-nm in single
  longitudinal mode with a diode-pumped {Pr}:{YLF} solid-state laser},\ }\href
  {https://doi.org/10.1016/j.optlastec.2019.03.025} {\bibfield  {journal}
  {\bibinfo  {journal} {Optics \& Laser Technology}\ }\textbf {\bibinfo
  {volume} {116}},\ \bibinfo {pages} {112} (\bibinfo {year}
  {2019})}\BibitemShut {NoStop}%
\bibitem [{\citenamefont {Yanagida}\ \emph {et~al.}(2009)\citenamefont
  {Yanagida}, \citenamefont {Yokota}, \citenamefont {Fujimoto}, \citenamefont
  {Yoshikawa}, \citenamefont {Kawaguchi}, \citenamefont {Ishizu}, \citenamefont
  {Fukuda}, \citenamefont {Suyama},\ and\ \citenamefont
  {Sarukura}}]{yanagida_crystal_2009}%
  \BibitemOpen
  \bibfield  {author} {\bibinfo {author} {\bibfnamefont {T.}~\bibnamefont
  {Yanagida}}, \bibinfo {author} {\bibfnamefont {Y.}~\bibnamefont {Yokota}},
  \bibinfo {author} {\bibfnamefont {Y.}~\bibnamefont {Fujimoto}}, \bibinfo
  {author} {\bibfnamefont {A.}~\bibnamefont {Yoshikawa}}, \bibinfo {author}
  {\bibfnamefont {N.}~\bibnamefont {Kawaguchi}}, \bibinfo {author}
  {\bibfnamefont {S.}~\bibnamefont {Ishizu}}, \bibinfo {author} {\bibfnamefont
  {K.}~\bibnamefont {Fukuda}}, \bibinfo {author} {\bibfnamefont
  {T.}~\bibnamefont {Suyama}},\ and\ \bibinfo {author} {\bibfnamefont
  {N.}~\bibnamefont {Sarukura}},\ }\bibfield  {title} {\bibinfo {title}
  {Crystal {Growth} and {Luminescence} {Properties} of {Pr}-{Doped} {LiYF}$_{\textrm{4}}$ and {LiCaAlF}$_{\textrm{6}}$},\ }\href
  {https://doi.org/10.1143/JJAP.48.085503} {\bibfield  {journal} {\bibinfo
  {journal} {Japanese Journal of Applied Physics}\ }\textbf {\bibinfo {volume}
  {48}},\ \bibinfo {pages} {085503} (\bibinfo {year} {2009})}\BibitemShut
  {NoStop}%
\bibitem [{\citenamefont {Kippenberg}\ \emph {et~al.}(2011)\citenamefont
  {Kippenberg}, \citenamefont {Holzwarth},\ and\ \citenamefont
  {Diddams}}]{kippenberg_microresonator-based_2011}%
  \BibitemOpen
  \bibfield  {author} {\bibinfo {author} {\bibfnamefont {T.~J.}\ \bibnamefont
  {Kippenberg}}, \bibinfo {author} {\bibfnamefont {R.}~\bibnamefont
  {Holzwarth}},\ and\ \bibinfo {author} {\bibfnamefont {S.~A.}\ \bibnamefont
  {Diddams}},\ }\bibfield  {title} {\bibinfo {title} {Microresonator-{Based}
  {Optical} {Frequency} {Combs}},\ }\href
  {https://doi.org/10.1126/science.1193968} {\bibfield  {journal} {\bibinfo
  {journal} {Science}\ }\textbf {\bibinfo {volume} {332}},\ \bibinfo {pages}
  {555} (\bibinfo {year} {2011})}\BibitemShut {NoStop}%
\bibitem [{\citenamefont {Barnes}\ and\ \citenamefont
  {Gettemy}(1980)}]{barnes_temperature_1980}%
  \BibitemOpen
  \bibfield  {author} {\bibinfo {author} {\bibfnamefont {N.~P.}\ \bibnamefont
  {Barnes}}\ and\ \bibinfo {author} {\bibfnamefont {D.~J.}\ \bibnamefont
  {Gettemy}},\ }\bibfield  {title} {\bibinfo {title} {Temperature variation of
  the refractive indices of yttrium lithium fluoride},\ }\href
  {https://doi.org/10.1364/JOSA.70.001244} {\bibfield  {journal} {\bibinfo
  {journal} {Journal of the Optical Society of America}\ }\textbf {\bibinfo
  {volume} {70}},\ \bibinfo {pages} {1244} (\bibinfo {year}
  {1980})}\BibitemShut {NoStop}%
\bibitem [{\citenamefont {Aggarwal}\ \emph {et~al.}(2005)\citenamefont
  {Aggarwal}, \citenamefont {Ripin}, \citenamefont {Ochoa},\ and\ \citenamefont
  {Fan}}]{aggarwal_measurement_2005}%
  \BibitemOpen
  \bibfield  {author} {\bibinfo {author} {\bibfnamefont {R.~L.}\ \bibnamefont
  {Aggarwal}}, \bibinfo {author} {\bibfnamefont {D.~J.}\ \bibnamefont {Ripin}},
  \bibinfo {author} {\bibfnamefont {J.~R.}\ \bibnamefont {Ochoa}},\ and\
  \bibinfo {author} {\bibfnamefont {T.~Y.}\ \bibnamefont {Fan}},\ }\bibfield
  {title} {\bibinfo {title} {Measurement of thermo-optic properties of
  {Y$_3$Al$_5$O$_{12}$}, {Lu$_3$Al$_5$O$_{12}$}, {YAIO$_3$}, {LiYF$_4$}, {LiLuF$_4$}, {BaY$_2$F$_8$}, {KGd}({WO$_4$})$_2$,
  and {KY}({WO$_4$})$_2$ laser crystals in the 80–{300K} temperature range},\
  }\href {https://doi.org/10.1063/1.2128696} {\bibfield  {journal} {\bibinfo
  {journal} {Journal of Applied Physics}\ }\textbf {\bibinfo {volume} {98}},\
  \bibinfo {pages} {103514} (\bibinfo {year} {2005})}\BibitemShut {NoStop}%
\bibitem [{\citenamefont {He}\ \emph {et~al.}(2009)\citenamefont {He},
  \citenamefont {Xiao}, \citenamefont {Zhu}, \citenamefont {Ozdemir},\ and\
  \citenamefont {Yang}}]{he2009OscillatoryThermalDynamics}%
  \BibitemOpen
  \bibfield  {author} {\bibinfo {author} {\bibfnamefont {L.}~\bibnamefont
  {He}}, \bibinfo {author} {\bibfnamefont {Y.-F.}\ \bibnamefont {Xiao}},
  \bibinfo {author} {\bibfnamefont {J.}~\bibnamefont {Zhu}}, \bibinfo {author}
  {\bibfnamefont {S.~K.}\ \bibnamefont {Ozdemir}},\ and\ \bibinfo {author}
  {\bibfnamefont {L.}~\bibnamefont {Yang}},\ }\bibfield  {title} {\bibinfo
  {title} {Oscillatory thermal dynamics in high-{{Q PDMS-coated}} silica
  toroidal microresonators},\ }\href@noop {} {\bibfield  {journal} {\bibinfo
  {journal} {Opt. Express, OE}\ }\textbf {\bibinfo {volume} {17}},\ \bibinfo
  {pages} {9571} (\bibinfo {year} {2009})}\BibitemShut {NoStop}%
\bibitem [{\citenamefont {Kobatake}\ \emph {et~al.}(2016)\citenamefont
  {Kobatake}, \citenamefont {Kato}, \citenamefont {Itobe}, \citenamefont
  {Nakagawa},\ and\ \citenamefont {Tanabe}}]{kobatake2016Thermal}%
  \BibitemOpen
  \bibfield  {author} {\bibinfo {author} {\bibfnamefont {T.}~\bibnamefont
  {Kobatake}}, \bibinfo {author} {\bibfnamefont {T.}~\bibnamefont {Kato}},
  \bibinfo {author} {\bibfnamefont {H.}~\bibnamefont {Itobe}}, \bibinfo
  {author} {\bibfnamefont {Y.}~\bibnamefont {Nakagawa}},\ and\ \bibinfo
  {author} {\bibfnamefont {T.}~\bibnamefont {Tanabe}},\ }\bibfield  {title}
  {\bibinfo {title} {Thermal {{Effects}} on {{Kerr Comb Generation}} in a
  {{CaF$_2$ Whispering-Gallery Mode Microcavity}}},\ }\href@noop {} {\bibfield
  {journal} {\bibinfo  {journal} {IEEE Photonics Journal}\ }\textbf {\bibinfo
  {volume} {8}},\ \bibinfo {pages} {1} (\bibinfo {year} {2016})}\BibitemShut
  {NoStop}%
\bibitem [{\citenamefont {Sedlmeir}(2016)}]{sedlmeir2016crystalline}%
  \BibitemOpen
  \bibfield  {author} {\bibinfo {author} {\bibfnamefont {F.}~\bibnamefont
  {Sedlmeir}},\ }\emph {\bibinfo {title} {Crystalline whispering gallery mode
  resonators}},\ \href@noop {} {Ph.D. thesis},\ \bibinfo  {school}
  {Friedrich-Alexander-Universit{\"a}t Erlangen-N{\"u}rnberg (FAU)} (\bibinfo
  {year} {2016})\BibitemShut {NoStop}%
\bibitem [{\citenamefont {Gorodetsky}\ and\ \citenamefont
  {Ilchenko}(1999)}]{gorodetsky_optical_1999}%
  \BibitemOpen
  \bibfield  {author} {\bibinfo {author} {\bibfnamefont {M.~L.}\ \bibnamefont
  {Gorodetsky}}\ and\ \bibinfo {author} {\bibfnamefont {V.~S.}\ \bibnamefont
  {Ilchenko}},\ }\bibfield  {title} {\bibinfo {title} {Optical microsphere
  resonators: optimal coupling to high-{Q} whispering-gallery modes},\ }\href
  {https://doi.org/10.1364/JOSAB.16.000147} {\bibfield  {journal} {\bibinfo
  {journal} {Journal of the Optical Society of America B}\ }\textbf {\bibinfo
  {volume} {16}},\ \bibinfo {pages} {147} (\bibinfo {year} {1999})}\BibitemShut
  {NoStop}%
\bibitem [{\citenamefont {Sedlmeir}\ \emph {et~al.}(2014)\citenamefont
  {Sedlmeir}, \citenamefont {Zeltner}, \citenamefont {Leuchs},\ and\
  \citenamefont {Schwefel}}]{sedlmeir_high-q_2014}%
  \BibitemOpen
  \bibfield  {author} {\bibinfo {author} {\bibfnamefont {F.}~\bibnamefont
  {Sedlmeir}}, \bibinfo {author} {\bibfnamefont {R.}~\bibnamefont {Zeltner}},
  \bibinfo {author} {\bibfnamefont {G.}~\bibnamefont {Leuchs}},\ and\ \bibinfo
  {author} {\bibfnamefont {H.~G.~L.}\ \bibnamefont {Schwefel}},\ }\bibfield
  {title} {\bibinfo {title} {High-$Q$ MgF$_2$ whispering gallery mode
  resonators for refractometric sensing in aqueous environment},\ }\href
  {https://doi.org/10.1364/OE.22.030934} {\bibfield  {journal} {\bibinfo
  {journal} {Optics Express}\ }\textbf {\bibinfo {volume} {22}},\ \bibinfo
  {pages} {30934} (\bibinfo {year} {2014})}\BibitemShut {NoStop}%
\bibitem [{\citenamefont {Gorodetsky}\ and\ \citenamefont
  {Ilchenko}(1994)}]{gorodetsky1994coupling}%
  \BibitemOpen
  \bibfield  {author} {\bibinfo {author} {\bibfnamefont {M.~L.}~\bibnamefont
  {Gorodetsky}}\ and\ \bibinfo {author} {\bibfnamefont {V.~S.}~\bibnamefont
  {Ilchenko}},\ }\bibfield  {title} {\bibinfo {title} {High-{{Q}} optical
  whispering-gallery microresonators: Precession approach for spherical mode
  analysis and emission patterns with prism couplers},\ }\href@noop {}
  {\bibfield  {journal} {\bibinfo  {journal} {Optics Communications}\ }\textbf
  {\bibinfo {volume} {113}},\ \bibinfo {pages} {133} (\bibinfo {year}
  {1994})}\BibitemShut {NoStop}%
\bibitem [{\citenamefont {Savchenkov}\ \emph {et~al.}(2014)\citenamefont
  {Savchenkov}, \citenamefont {Liang}, \citenamefont {Ilchenko}, \citenamefont
  {Dale}, \citenamefont {Savchenkova}, \citenamefont {Matsko}, \citenamefont
  {Seidel},\ and\ \citenamefont {Maleki}}]{Savchenkov14Esens}%
  \BibitemOpen
  \bibfield  {author} {\bibinfo {author} {\bibfnamefont {A.~A.}\ \bibnamefont
  {Savchenkov}}, \bibinfo {author} {\bibfnamefont {W.}~\bibnamefont {Liang}},
  \bibinfo {author} {\bibfnamefont {V.~S.}\ \bibnamefont {Ilchenko}}, \bibinfo
  {author} {\bibfnamefont {E.}~\bibnamefont {Dale}}, \bibinfo {author}
  {\bibfnamefont {E.~A.}\ \bibnamefont {Savchenkova}}, \bibinfo {author}
  {\bibfnamefont {A.~B.}\ \bibnamefont {Matsko}}, \bibinfo {author}
  {\bibfnamefont {D.}~\bibnamefont {Seidel}},\ and\ \bibinfo {author}
  {\bibfnamefont {L.}~\bibnamefont {Maleki}},\ }\bibfield  {title} {\bibinfo
  {title} {Photonic e-field sensor},\ }\href@noop {} {\bibfield  {journal}
  {\bibinfo  {journal} {AIP Advances}\ }\textbf {\bibinfo {volume} {4}},\
  \bibinfo {pages} {122901} (\bibinfo {year} {2014})}\BibitemShut {NoStop}%
\bibitem [{\citenamefont {Savchenkov}\ \emph {et~al.}(2018)\citenamefont
  {Savchenkov}, \citenamefont {Borri}, \citenamefont {{Siciliani de Cumis}},
  \citenamefont {Matsko}, \citenamefont {De~Natale},\ and\ \citenamefont
  {Maleki}}]{savchenkov2018Q}%
  \BibitemOpen
  \bibfield  {author} {\bibinfo {author} {\bibfnamefont {A.~A.}\ \bibnamefont
  {Savchenkov}}, \bibinfo {author} {\bibfnamefont {S.}~\bibnamefont {Borri}},
  \bibinfo {author} {\bibfnamefont {M.}~\bibnamefont {{Siciliani de Cumis}}},
  \bibinfo {author} {\bibfnamefont {A.~B.}\ \bibnamefont {Matsko}}, \bibinfo
  {author} {\bibfnamefont {P.}~\bibnamefont {De~Natale}},\ and\ \bibinfo
  {author} {\bibfnamefont {L.}~\bibnamefont {Maleki}},\ }\bibfield  {title}
  {\bibinfo {title} {Modeling and measuring the quality factor of whispering
  gallery mode resonators},\ }\href@noop {} {\bibfield  {journal} {\bibinfo
  {journal} {Applied Physics B}\ }\textbf {\bibinfo {volume} {124}},\ \bibinfo
  {pages} {171} (\bibinfo {year} {2018})}\BibitemShut {NoStop}%
\bibitem [{\citenamefont {Foreman}\ \emph {et~al.}(2016)\citenamefont
  {Foreman}, \citenamefont {Sedlmeir}, \citenamefont {Schwefel},\ and\
  \citenamefont {Leuchs}}]{foreman_dielectric_2016}%
  \BibitemOpen
  \bibfield  {author} {\bibinfo {author} {\bibfnamefont {M.~R.}\ \bibnamefont
  {Foreman}}, \bibinfo {author} {\bibfnamefont {F.}~\bibnamefont {Sedlmeir}},
  \bibinfo {author} {\bibfnamefont {H.~G.~L.}\ \bibnamefont {Schwefel}},\ and\
  \bibinfo {author} {\bibfnamefont {G.}~\bibnamefont {Leuchs}},\ }\bibfield
  {title} {\bibinfo {title} {Dielectric tuning and coupling of whispering
  gallery modes using an anisotropic prism},\ }\href
  {https://doi.org/10.1364/JOSAB.33.002177} {\bibfield  {journal} {\bibinfo
  {journal} {JOSA B}\ }\textbf {\bibinfo {volume} {33}},\ \bibinfo {pages}
  {2177} (\bibinfo {year} {2016})}\BibitemShut {NoStop}%
\end{thebibliography}
\end{document}


\title{Distance calibration via Newton's rings in yttrium lithium fluoride whispering gallery mode resonators: supplemental document}

\maketitle

\section{Orientation of YLF crystal} \label{sec:cryts}
\begin{figure}[hbt]
\centering
	\includegraphics[width=0.5\linewidth]{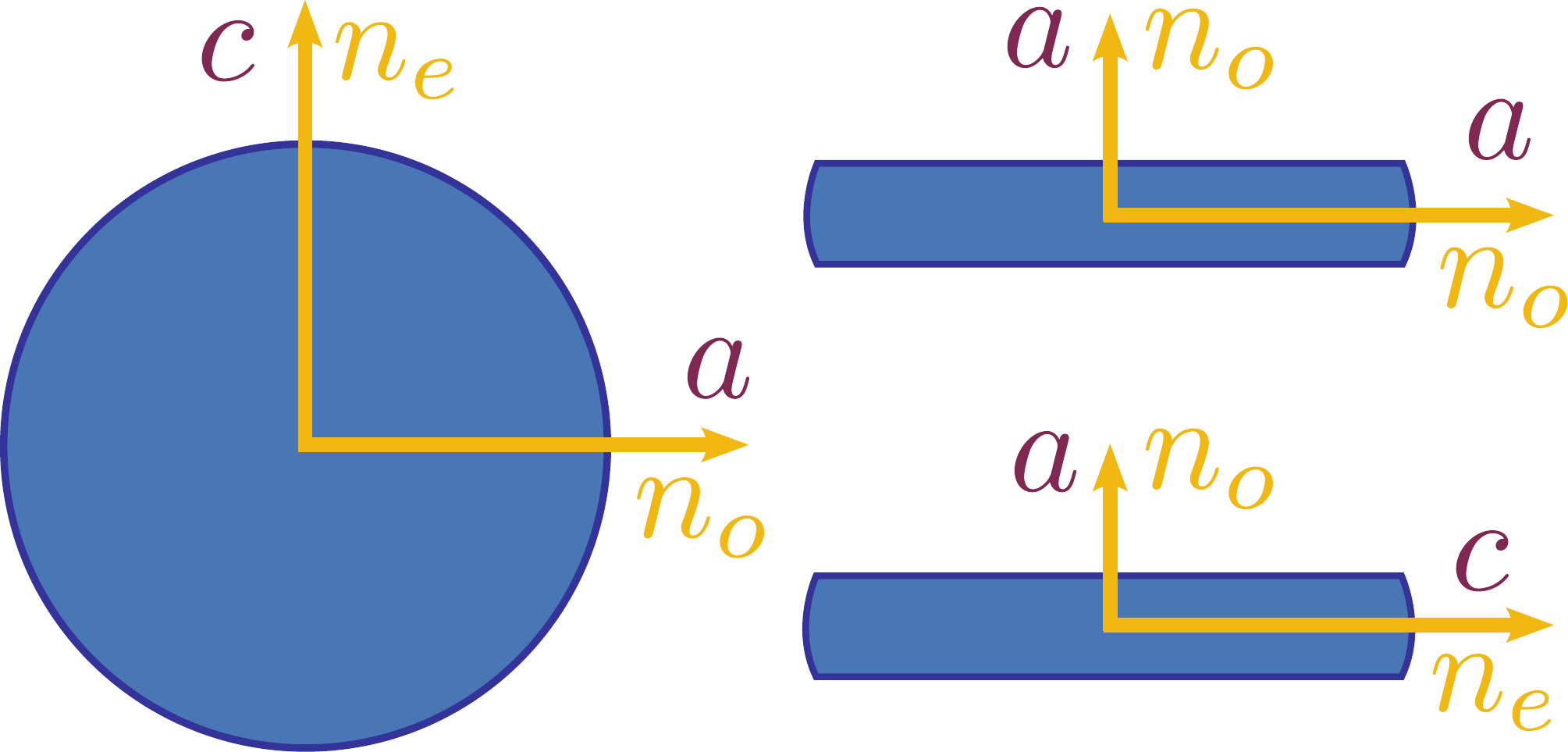}
	\caption{2D schematic indicating the refractive indices along each axis. $n_o$ is the ordinary and $n_e$ is the extraordinary refractive index.}
	\label{fig:ResoIndex}
\end{figure}

Yttrium lithium fluoride is a uniaxial material~\cite{pollak_yttrium_1980}, and has a Scheelite-like structure with the symmetry space group $I4_1/a$~\cite{grzechnik_scheelite_2002}. The crystal axes of our YLF crystal are pictured in Fig.~\ref{fig:ResoIndex}, which we refer to as `$a$-cut'.

\section{Beam alignment through a trapezoidal prism}\label{BeamAl}
\begin{figure}
 
 \begin{subfigure}{.475\textwidth}
 	\centering
 	\includegraphics[width=\textwidth]{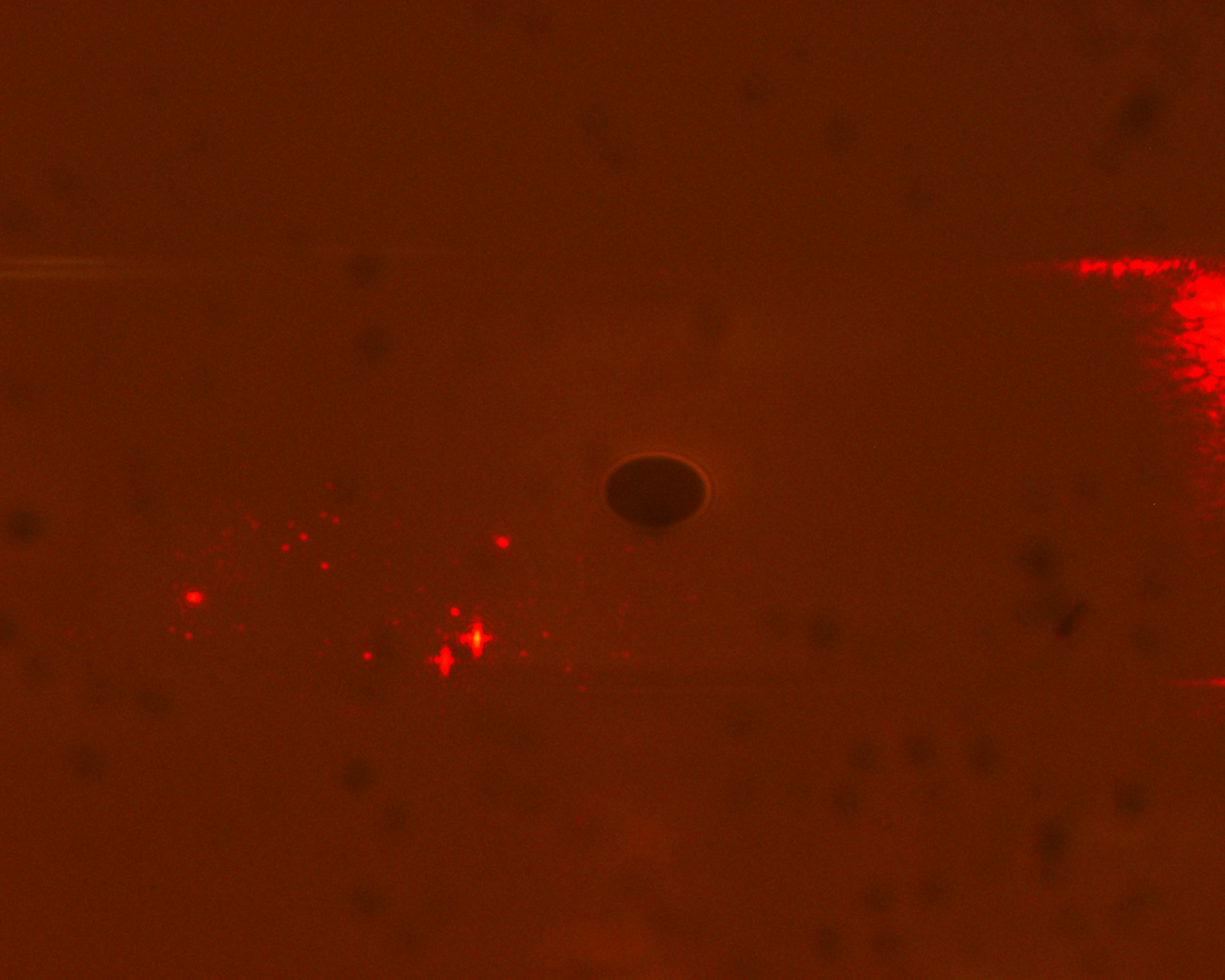}
 	\caption{Initial state with unfocused beam, misaligned from the Newton's rings (seen as dark spot in the middle). The red guiding laser induces scattering from dust particles in the southwest region of the Newton's rings.}
 	\label{fig:coup1}
 \end{subfigure}
\hfill
 \begin{subfigure}{.475\textwidth}
 	 	\centering
	\includegraphics[width=\textwidth]{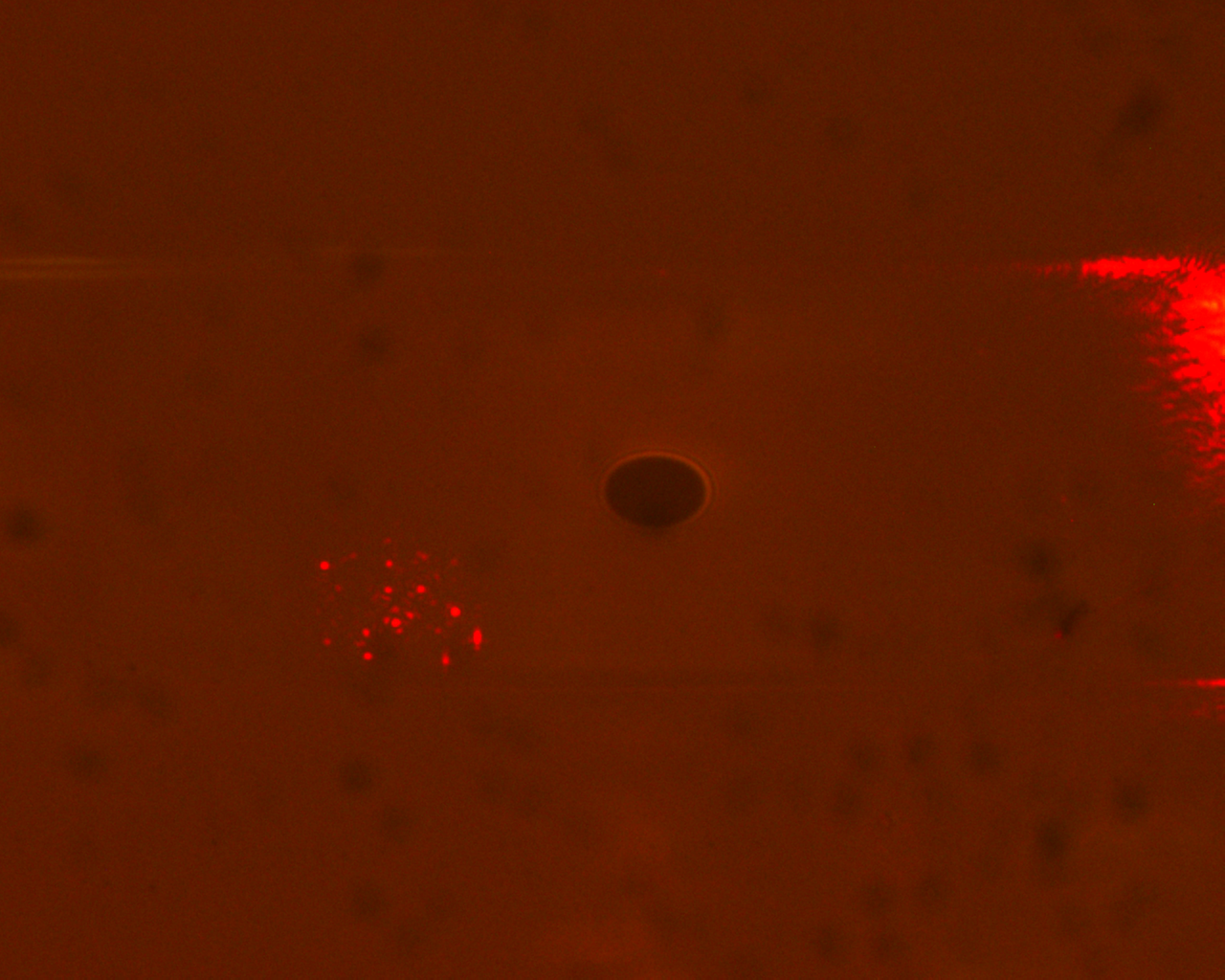}
	\caption{Adjusting the gap between the pigtailed ferrule and GRIN lens to achieve as narrow a beam as possible.\vspace{2\baselineskip}}
	\label{fig:coup2}
\end{subfigure}

\begin{subfigure}{.475\textwidth}
 	 	\centering
	\includegraphics[width=\textwidth]{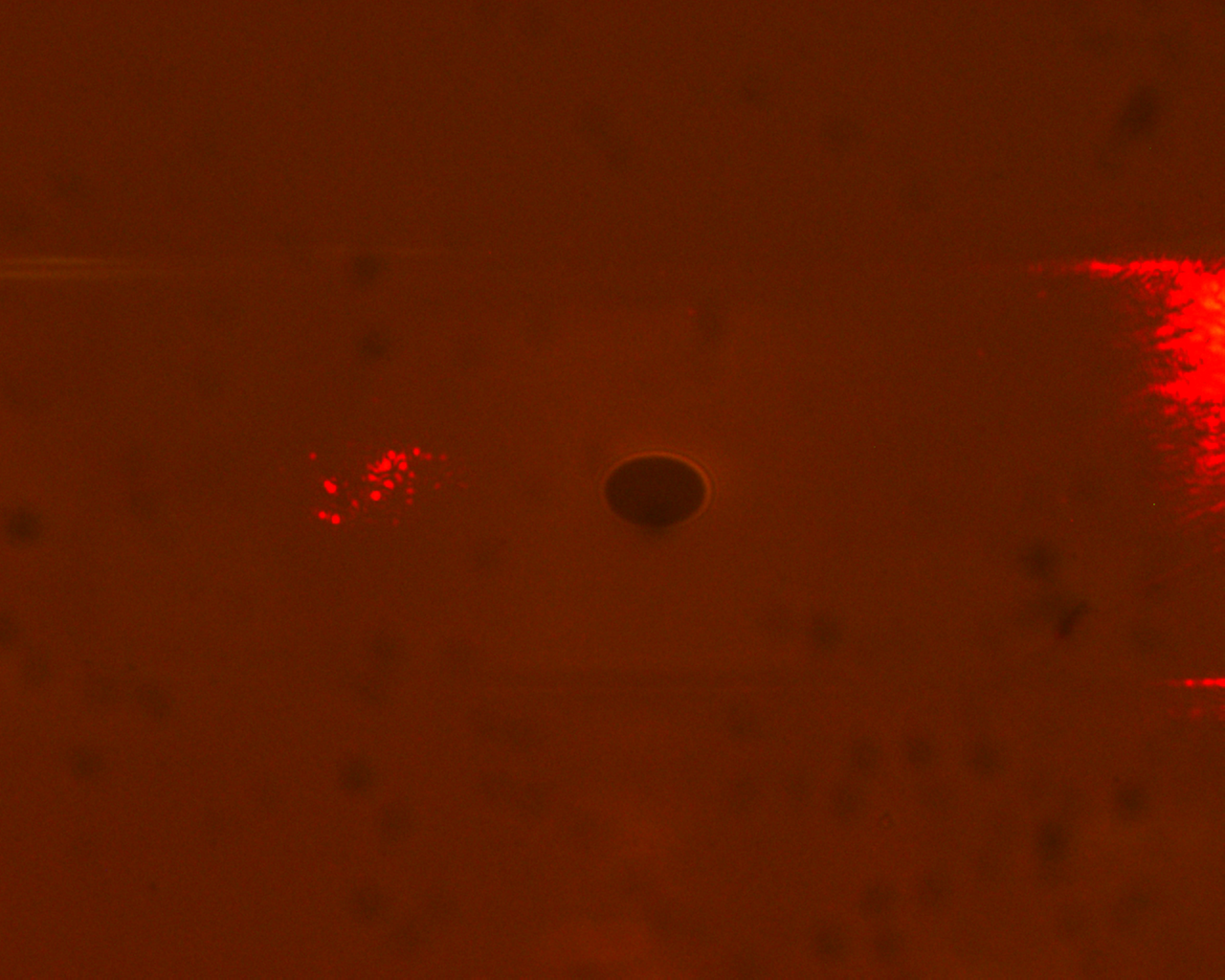}
	\caption{Adjustment of the position of the GRIN lens northward to align with the Newton's rings.\vspace{2\baselineskip}}
	\label{fig:coup3}
\end{subfigure}
\hfill
 \begin{subfigure}{.475\textwidth}
 	 	\centering
	\includegraphics[width=\textwidth]{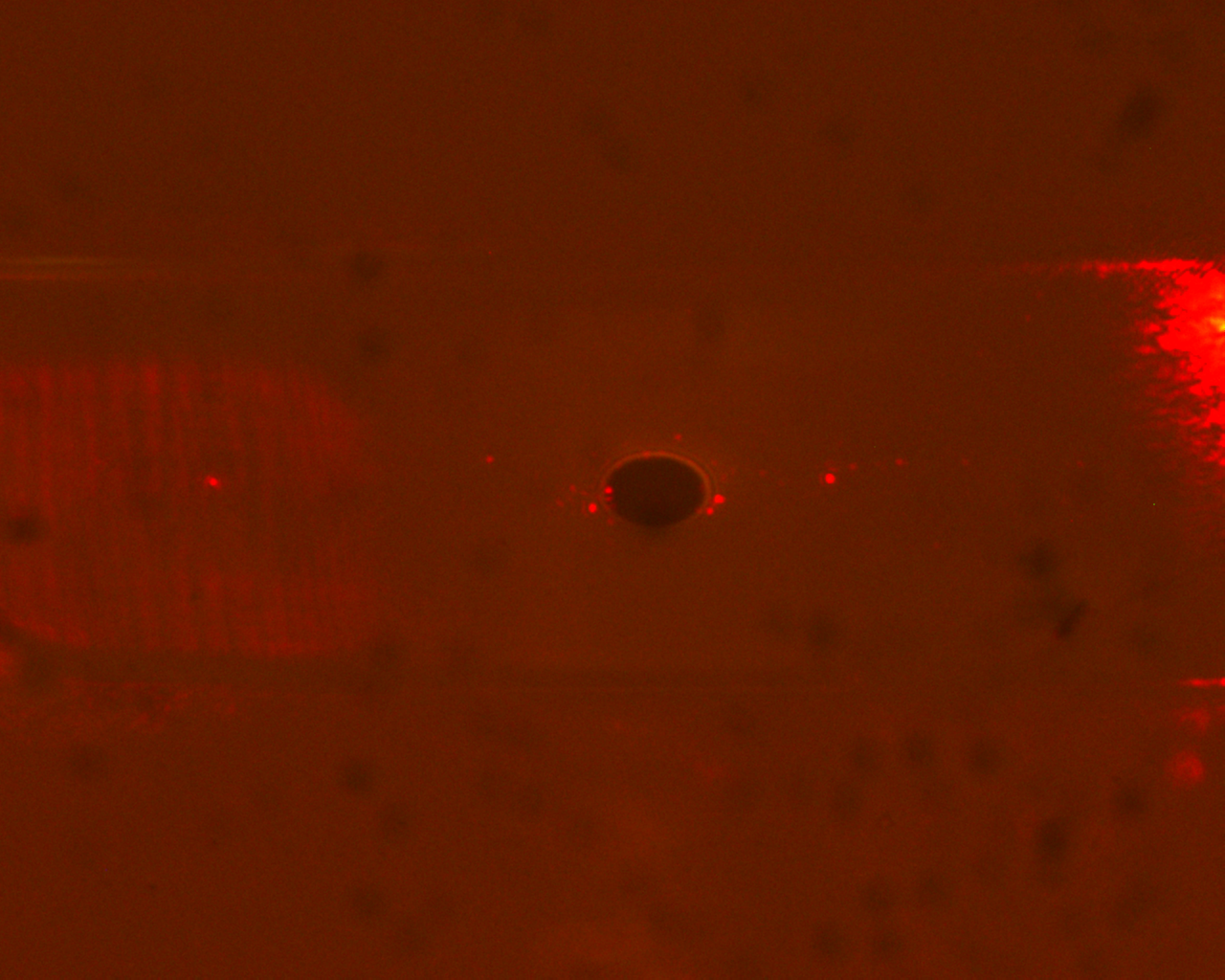}
	\caption{Adjustment of the position of the GRIN lens eastward to align with the Newton's rings and achieve overlap, resulting in coupling which can be seen as the glow in the western half of the image.}
	\label{fig:coup4}
\end{subfigure}
\caption{A series of adjustments made to accurately couple light into the resonator, viewed through a microscope magnified into the central rim of our \SI{2.94}{\mm} diameter resonator.}
\label{fig:coup}
\end{figure}

A camera was positioned to look and focus into the resonator from behind the prism, through the shorter base of the prism. Newton's rings were formed when the prism was brought in contact with the resonator, this point was used as reference as we switched on our red guiding laser and saw broad laser reflections displaced away from the rings, see Figure~\ref{fig:coup}. We configured the focus of the graded index (GRIN) lens noting the smaller area of reflections as a result, Figure~\ref{fig:coup2}. We continued to align the laser by adjusting a three-axis translation stage -- atop which the GRIN lens is mounted -- which corresponded to displacements of the laser reflections on the resonator respectively. By overlapping these laser reflections with the Newton's rings we achieved coupling into the resonator (Figure~\ref{fig:coup4}). 

\section{Explanation of fitting parameters}

The equation to which we fit the ring-averaged reflected-light intensity profile is repeated below: 
\begin{equation}
\label{eq:sinfit}
    I = p_1\sin^2{(p_2x^2 + \phi)}e^{-p_3x^2} +p_4+p_5x+p_6x^2.
\end{equation}
As stated in the main text, the phase $\phi$ gives the distance between resonator and prism (once ambiguity of zero phase is removed):
\begin{equation}
\label{eq:v2d}
    d  = \frac{\phi\lambda_f}{2\pi}.
\end{equation}
$p_1$ gives the visibility of the Newton's rings at their center.
Away from the center of the rings, there is an additional gap between the resonator and prism due to the curvature of the resonator. In the area we look at, this additional gap is quadratic with number of pixels away from the center, hence the $p_2x^2$ term.
The fringe visibility will decay away from the center due to a loss of coherence between the two reflections. In particular, the relative phase of the two reflections is increasingly ambiguous due to uncertainty in the precise path the light is reflected along as well as the bandwidth of the band-pass filter used. We assume the visibility decays exponentially with the distance between resonator and prism, i.e.\ proportional to $\exp(-p_3x^2)$.
Destructive interference will not be complete even if the resonator and prism touch. In this limiting case there will still be some light reflected at the prism--resonator interface. This necessitates a constant background term $p_4$. $p_5$ and $p_6$ account for both uneven illumination in some cases as well as the changing background reflection about which the rings oscillate.\newpage 


%